
%
%
%
%
\input amstex
\documentstyle{amsppt}

\document
\magnification=\magstep1


\baselineskip=13pt
\let\<=\langle
\let\>=\rangle
\global\def\ssectitle#1\par{\bigbreak\medskip
  \leftline{\typc #1}
  \nobreak\bigskip\vskip-\parskip
  \message{#1}
  \noindent}
\def\today{\ifcase\month\or
  January\or February\or March\or April\or May\or June\or
  July\or August\or September\or October\or November\or December\fi
  \space\number\day, \number\year}

\def\ss{\goth S}
\def\bb{\goth B}

\def\AA{\Cal A}

\def\cc{\Cal C}
\def\xx{\bold x}

\def\00{\bold 0}
\def\hn{H^{(n)}}
\def\wn{W^{(n)}}
\def\bn{\bb^{(n)}}
\def\hm{H^{(m)}}
\def\wm{W^{(m)}}
\def\bm{\bb^{(m)}}

\font\typc=cmbx10 scaled \magstep1  
\font\typg=cmcsc10 scaled \magstep2 
\font\typy=cmr10  
\font\type=cmbx10 scaled \magstep3  

\hrule
\medskip
\rightline{PAR-LPTHE 93/33}
\rightline{hep-th/9306053}
\medskip
\hrule
\bigskip
\bigskip
\bigskip

\centerline{\typg Sergey Fomin}

\medskip

{\typy
\settabs 2 \columns
\+  \quad\ \qquad Department of Mathematics,
& \quad\ \qquad \quad Theory of Algorithms Laboratory
\cr
\+ \quad\  Massachusetts Institute of Technology
& \quad\ \qquad \quad \quad \quad \quad
SPIIRAN, Russia \cr
}

\bigskip
\bigskip

\centerline{\typg Anatol N. Kirillov}

\medskip

{\typy
\settabs 2 \columns
\+  \qquad\ \qquad \qquad L.P.T.H.E.

& \quad\qquad \qquad  Steklov Mathematical Institute
\cr
\+ \qquad\ \qquad  Universit\'e Paris VI
& \quad\quad \qquad \quad \quad \quad
St.~Petersburg, Russia \cr
}

\bigskip

\bigskip
\bigskip
\centerline{\type{Combinatorial B$_{\tenrm{n}}$-analogues}}
\bigskip
\centerline{\type{of Schubert polynomials}}
\bigskip
\endtopmatter

\ssectitle{Abstract}

Combinatorial $B_n$-analogues of Schubert polynomials and corresponding
symmetric functions are constructed from an exponential solution
of the $B_n$-Yang-Baxter equation that involves the nilCoxeter algebra
of the hyperoctahedral group.

\ssectitle{1. Introduction}

Schubert polynomials of Lascoux and Sch\"utzenberger
(see \cite{L, M} and the literature therein)
can be introduced and studied from different points of view.
In this paper, we adapt to the $B_n$ case
the approach to the theory of Schubert polynomials
that was developed in \cite{FS, FK1} (see also \cite{FK2, FK3});
this approach is based on an exponential solution
of the Yang-Baxter equation that is related to the nilCoxeter algebra
of the symmetric ($A_n$ case) or the hyperoctahedral ($B_n$ case) group.
On this way, we introduce certain polynomials
$\bn_w$ which can be viewed
as candidates for $B_n$-Schubert polynomials.
The algebraic-geometry sense of these polynomials is at least obscure.
However, they prove to have nice combinatorial properties
which are similar to those of ordinary Schubert polynomials.
In particular:

(i) they satisfy similar {\it recurrences with divided differences}
(with a single exception that corresponds to the "special" generator $s_0$
of the hyperoctahedral group $W^{(n)}$ with generators $s_0$, \dots,
$s_{n-1}$);

(ii) they have direct combinatorial interpretation in terms of
``$B_n$-braids''; it means, in particular, that they have
{\it nonnegative coefficients};

(iii) this interpretation can be restated in terms of {\it reduced
decompositions} of $w$ and {\it ``compatible'' sequences};

(iv) the defining expression in the nilCoxeter algebra has $n^2$
factors; these factors are in natural bijection with the entries
of a standard reduced decomposition of $w_0$,
{\it the element of maximal length} in $W^{(n)}$;

(v) polynomials $\bn_w$ have a {\it stability} property,
in the sense that the coefficient of any monomial in $\bn_w$ gets
fixed when $n$ is sufficiently large;

(vi) we introduce {\it symmetric functions} $H_w$
which have a similar combinatorial interpretation and can be obtained
as a certain {\it limit} of $\bn_w$'s; these  functions generalize
Schur $P$-functions.

This paper is organized as follows.
Section 2 contains a straightforward adaptation for the $B_n$ case
of the main ``geometric'' construction used in \cite{FK1};
the role of the Yang-Baxter equation (YBE) is briefly explained.
(At this point, an acquaintance with our ``$A_n$''-paper \cite{FK1}
would be very helpful.)
In Section 3, some exponential solutions of $B_n$-YBE are listed (we refer
to \cite{FK3} for details).
Section 4 introduces $B_n$-symmetric functions which one can
associate with any such solution.
Generalized Schubert expressions for the $B_n$ case are
suggested in Section 5.
For the nilCoxeter algebra solution,
these expressions give rise to polynomials $\bn_w$
which are studied in Section 6.
Section 7 contains the full list of $\bn_w$~'s for $n\leq 3$.
Section 8 is devoted to $B_n$-analogues of stable Schubert
polynomials.

{\smc Acknowledgements.} The authours are grateful to Richard Stanley
for helpful comments.

\ssectitle{2. Generalized configurations and the Yang-Baxter equation}

The notion of {\it generalized configuration} was introduced in \cite{FK1}.
It is a configuration of contiguous lines which cross a given
vertical strip from left to right;
each line is subdivided into ``segments'';
each segment has an associated variable.
A configuration is assumed to be generic in the following sense:
(i) no three lines intersect at the same point;
(ii) no two lines intersect at an endpoint of any segment;
(ii) no two intersection points lie on the same vertical line.

In the $B_n$ case, this notion assumes an additional flavour.
Namely, configurations are contained in a {\it semi-strip}
bounded from below by a {\it bottom mirror} (cf. \cite{Ch}).
The lines of a configuration are allowed to touch the bottom;
whenever it happens, the associated variable changes its sign.
An example of a $B_n$-configuration is given on Figure 1.

We are particularly interested in {\it intersection points}
and {\it points of reflection}.
Each intersection point has a {\it level number} which indicates
how many lines are there below this point
(the point itself contributes 1).
For example, the intersection points on Figure 1 have level numbers
(from left to right) 1, 1, 2, and 1.
The level number of a point of reflection is 0 by definition.

Let $\cc$ be a configuration of the described type.
Let us order its intersection and reflection points altogether
from left to right; then write down their level numbers.
The resulting sequence of integers $a(\cc)=a_1a_2\dots$ is called
a {\it word associated} with $\cc$.
In our running example, $a(\cc)=101201$.
Now it is time to bring the variables into the picture.
Assume that $\AA$ is an associative algebra and
$\{h_i(x)\ :\ i=0,1,\dots\}$ is a family of elements of $\AA$
which depend on a formal variable $x$
(we always assume that the main field contains all participating
formal variables).
Then the {\it associated expression} for a configuration $\cc$
is
$$ \Phi(\cc) = h_{a_1}(z_1)\ h_{a_2}(z_2)\ \dots $$
where, as before, $a_1a_2\dots$ is an associated word
and $z_i$ is one of the following: if $a_i=0$, then $z_i$
is the variable related to the corresponding point of reflection
(to the left of it); if $a_i>0$, then $z_i=x_i-y_i$ where
$x_i$ and $y_i$ are the variables for the segments intersecting at the
corresponding points, $x_i$ being associated with the segment
which is above to the left of this point.
In the example of Figure 1,
$$ \Phi(\cc) = \Phi(\cc; x, y, z, u, v) =
h_1(y-u) h_0(y) h_1(v+y) h_2(x+y) h_0(v) h_1(x+v)\ . $$
Informally, the variables associated with the segments
are their ``slopes''; an argument of each factor in $ \Phi(\cc)$
is the corresponding ``angle of intersection''.

The {\it Yang-Baxter equations} (see, e.g., \cite{Ch} and references therein)
are certain conditions on $h_i(x)$'s which allow to transform
configurations without changing their associated expression.
In the case under consideration (i.e., in the $B_n$ case),
the YBE are
$$ h_i(x)h_j(y)=h_j(y)h_i(x) \quad \text{if} \quad |i-j|\geq 2 \ ; \tag{2.1}$$
$$ h_i(x)h_{i+1}(x+y)h_i(y) = h_{i+1}(y)h_i(x+y)h_{i+1}(x)
\quad \text{if}\quad i\geq 1\ ; \tag{2.2} $$
$$ h_1(x-y)h_0(x)h_1(x+y)h_0(y) =  h_0(y)h_1(x+y)h_0(x)h_1(x-y)
\ . \tag{2.3} $$
Each of these equations has its pictorial interpretation;
see Figure 2.

Following \cite{FK1}, we introduce the additional condition
$$ h_i(x)h_i(y) = h_i(x+y)\ ,\quad h_i(0)=1 \quad \text{if}\quad i\geq 1\
\tag{2.4} $$
(cf. Figure 3). Conditions (2.1)-(2.4) have various nice
implications. To derive them, one can use ``geometric'' techniques
instead of cumbersome algebraic computations.
Informally, algebraic identities can be proved by moving lines
according to the rules of Figures 2-3.

\ssectitle{3. Examples of solutions}

There is a natural way to construct solutions of the equations
(2.1)-(2.4). Assume $\AA$ is a {\it local} algebra with generators
$u_0$, $u_1$, $u_2$, \dots which means that
$$ u_iu_j = u_ju_i \quad \text{if} \quad |i-j|\geq 2 \ . \tag{3.1} $$
Define $h_i(x)$ by
$$ h_i(x)=\exp(xu_i)\ ; \tag{3.2} $$
then (2.1) and (2.4) are automatically guaranteed,
and we only need the Yang-Baxter equations (2.2)-(2.3)
to be satisfied. Rewrite (2.2)-(2.3) as
$$ e^{xu_i} e^{(x+y)u_{i+1}} e^{yu_i} =
e^{yu_{i+1}} e^{(x+y)u_i} e^{xu_{i+1}} \tag{3.3} $$
and
$$ [e^{xu_0}e^{xu_1}e^{xu_0},e^{yu_0}e^{yu_1}e^{yu_0}] = 0 \ . \tag{3.4} $$
These equations were studied in \cite{FK3} where the following solutions
were suggested.

\proclaim{3.1 Example} The nilCoxeter algebra of the
hyperoctahedral group.
\rm
This is the algebra defined by
$$
\align
& u_iu_j=u_ju_i\ , |i-j|\geq 2; \\
& u_i^2=0 \ ; \\
& u_iu_{i+1}u_i=u_{i+1}u_iu_{i+1}\ , i\geq 1\ ; \\
& u_0u_1u_0u_1=u_1u_0u_1u_0\ .
\endalign
$$
\endproclaim

\proclaim{3.2 Example} Universal enveloping algebra of $U_+(so(2n+1))$.
\rm
This algebra can be defined as the local algebra with generators
$u_1$, $u_2$, \dots
subject to Serre relations
$$
\align
& [u_i,[u_i,u_{i\pm 1}]]=0 \ , i\geq 1\ ;\\
& [u_0,[u_0,u_1]]=0 \ ; \\
& [u_1,[u_1,[u_1,u_0]]]=0 \ . \\
\endalign
$$
\endproclaim

Example 3.1 will be the main one in this paper.
Note that the nilCoxeter algebra can be alternatively defined
as the algebra spanned by
the elements of a Coxeter group, with the multiplication rule
$$ w\cdot v = \cases \text{usual product\ \ $wv$\ \ if\ \ $l(w)+l(v)=l(wv)$} \\
		0\ ,\ \text{otherwise}
	\endcases
$$
where $l(w)$ is the length of $w$
(the minimal number of generators whose product is $w$).

\ssectitle{4. Symmetric expressions}

By analogy with \cite{FK1}, we will show now that the basic relations
(2.1)-(2.4) (or (3.1)-(3.4)) imply that certain configurations
produce {\it symmetric} expressions in corresponding variables.
In what follows we assume that (2.1)-(2.4) are satisfied.

\proclaim{4.1 Theorem}
For the configuration $\cc$ of Figure 4,
$$ \Phi(\cc; x, y) = \Phi(\cc; y, x)\ . $$
In other words, $\Phi(\cc)$ is symmetric in $x$ and $y$.
\endproclaim

This statement has the following obvious reformulation.

\proclaim{4.2 Proposition}
Let
$$ B(x) = h_{n-1}(x) \cdot\cdot\cdot h_1(x) h_0(x) h_1(x) \cdot\cdot\cdot
h_{n-1}(x)\ . \tag{4.1} $$
Then $B(x)$ and $B(y)$ commute.
\endproclaim

{\bf Special case:} $n=2$. Then
$B(x)=h_1(x) h_0(x) h_1(x)$.
Now use (2.3) and (2.4)  to show that
$$
\align
B(x)B(y) & = h_1(x) h_0(x) h_1(x) h_1(y) h_0(y) h_1(y) \\
& = h_1(y) h_1(x-y) h_0(x) h_1(x+y) h_0(y) h_1(y) \\
& = h_1(y) h_0(y) h_1(x+y) h_0(x) h_1(x-y) h_1(y) \\
& = h_1(y) h_0(y) h_1(y) h_1(x) h_0(x) h_1(x) \\
& = B(y)B(x)\ .
\endalign
$$

The same proof can be performed in the language of configurations
--- see Figure 5. Moreover, the geometric proof has the advantage
of being easily adjustable for the general case of an arbitrary $n$.

{\bf Proof of Theorem 4.1} (and Proposition 4.2).
Same transformations as in Figure 5,
with additional horizontal lines added near the bottom. \ \qed

Proposition 4.2 implies that the expression
$$ \hn(x_1, x_2,\dots) = B(x_1)B(x_2) \cdot\cdot\cdot \tag{4.2} $$
is symmetric in $x_i$'s.
Therefore one can construct examples of symmetric functions
by taking any solution of (2.1)-(2.4), then any representation
of the corresponding algebra $\AA$, and applying the operator
representing $\hn(x_1, x_2,\dots)$ to any vector.

Surprisingly enough, the expression $\hn(x_1, x_2,\dots)$
can be alternatively defined by a quite different configuration.

\proclaim{4.3 Theorem}
Let $\cc$ be the configuration defined by Figure 6. Then
$$ \Phi(\cc; x_1, \dots, x_n) = \hn(x_1, \dots, x_n)\ . $$
\endproclaim

{\bf Proof.} See Figure 7. \ \qed

\noindent {\bf Remark.} If the number of generators is $m>n$,
then, to get $H^{(m)}(x_1, \dots, x_n)$, one only needs to
add $m-n$ horizontal lines near the bottom mirror
to the configuration of Figure~6.

Theorem 4.3 allows us to relate the $B_n$- and $A_n$-constructions
to each other. Note that the configuration of Figure 6
coincides with one of \cite{FK1, Figure 14} up to renumbering
the variables $x_i$ in the opposite order (this is not essential
since the expression is symmetric in $x_i$'s), setting $y_i=-x_i$,
and dropping (sorry) the bottom mirror. Since Figure 14
of \cite{FK1} defines the ordinary (i.e, $S_n$-) double stable Schubert
expression $G(x_1, \dots, x_n;y_1, \dots, y_n)$, the above observation
has the following precise formulation.

\proclaim{4.4 Theorem}
Let $\{h_i(x)\ :\ i=1,\dots,n-1\}$ be any solution of {\rm (2.1)},
{\rm (2.2)}, and {\rm (2.4)}; in other words, let $\{h_i(x)\}$ be an
exponential solution of the $A_{n-1}$-YBE.
Define $h_0(x)=1$. Then {\rm (2.3)} obviously holds and so $\hn$
is well-defined. Moreover, in this case
$$ \hn(x_1, \dots, x_n) = G(x_1, \dots, x_n; -x_1, \dots, -x_n) $$
where $G(\dots)$ is the double stable Schubert expression
(see \cite{FK1}).
\endproclaim

In the special case of the nilCoxeter solution of Example 3.1
we obtain $B_n$-analogues of Stanley's symmetric functions \cite{S}
(stable Schubert polynomials).
These functions are studied in Section 8.

\ssectitle{5. Schubert expressions}

Define the $B_n$-analogue of the generalized Schubert expression by
$$ \bn(x_1,\dots,x_n)=\hn(x_1,\dots,x_n)\ss(-x_1,\dots,-x_{n-1})
\tag{5.1} $$
where $\ss(x_1,\dots,x_{n-1})$ is the $A_{n-1}$-Schubert expression
as defined in \cite{FS, FK1}. In other words,
$$ \bn(x_1,\dots,x_n)=B(x_1)\cdot\cdot\cdot B(x_n)
A_1(-x_1)\cdot\cdot\cdot A_{n-1}(-x_{n-1})
\tag{5.2} $$
where
$$ A_i(x) = h_{n-1}h_{n-2}\cdot\cdot\cdot h_i(x)
\tag{5.3} $$
(recall that $B(x)$ is defined by (4.1)).

The formula (5.2) can be simplified.

\proclaim{5.1 Theorem}
$ \bn(x_1,\dots,x_n)$ is equal to the expression defined by Figure 8.
In other words,
$$ \bn(x_1,\dots,x_n)=\ss(x_n,x_{n-1},\dots,x_2)\
\prod_{i=0}^{n-1}\Biggl(h_0(x_{n-i})\
\prod_{j=1}^{n-i-1} h_j(x_{n-i-j}+x_{n-i}) \Biggr)
\tag{5.4} $$
where, as before,
$\ss(x_n,x_{n-1},\dots,x_2)=A_1(x_n)\cdot\cdot\cdot A_{n-1}(x_2)$
and in the products $\prod\cdot\cdot\cdot$ the factors are multiplied
left-to-right, according to the increase of $i$ and $j$,
respectively.
\endproclaim

Note that the total number of factors $h_{\dots}(\dots)$
in (5.4) is $n^2$, the length of the longest element $w_0$ of the
hyperoctahedral group $W^{(n)}$ with $n$ generators.
Moreover, it can be immediately seen from Figure 8 that
these factors are in a natural order-respecting
bijection with the entries of the standard reduced decomposition
of $w_0$:
$$ n-1,\ n-2,\ \dots,\ 2,\ 1,\
n-1,\ n-2,\ \dots,\ 2,\ 1,\ \
\dots\dots\ ,\
n-1,\ n-2,\ \dots,\ 2,\ 1\ . $$

\proclaim{5.2 Examples} \rm
$$
\align
n=1 \qquad & \bb^{(1)}(x_1)=h_0(x_1) \\
n=2 \qquad & \bb^{(2)}(x_1,x_2)=
h_1(x_2) h_0(x_2) h_1(x_1+x_2) h_0(x_1)\\
n=3 \qquad & \bb^{(3)}(x_1,x_2,x_3) \\
& =
h_2(x_3) h_1(x_3) h_2(x_2) h_0(x_3)
h_1(x_2+x_3) h_2(x_1+x_3) h_0(x_2) h_1(x_1+x_2) h_0(x_1)
\endalign
$$
\endproclaim

{\bf Proof of Theorem 5.1.}
Let $\Phi_6$ and $\Phi_8$ be the expressions defined by configurations
of Figures 6 and 8, respectively. Then
$$ \Phi_6 = \Phi_8 \tilde A_{n-1}(x_{n-1}) \cdot\cdot\cdot
\tilde A_2(x_2) \tilde A_1(x_1) $$
where
$$ \tilde A_i(x) = h_i(x)h_{i+1}(x)\cdot\cdot\cdot h_{n-1}(x) \ . $$
Since (2.4) implies that $(\tilde A_i(x))^{-1}=A_i(-x)$,
it follows from Theorem 4.3 that
$$
\align
\Phi_8 &= \Phi_6 A_1(-x_1)A_2(-x_2)\cdot\cdot\cdot A_{n-1}(-x_{n-1}) \\
& = \hn(x_1,\dots,x_n)\ss(-x_1,\dots,-x_{n-1}) \\
& = \bn(x_1,\dots,x_n)\ .\ \qed
\endalign
$$

\ssectitle{6. B$_n$-analogues of Schubert polynomials}

In the rest of this paper we study the main example of solution
of (2.1)-(2.4), namely, the one related to the nilCoxeter algebra
of the hyperoctahedral group (Example 3.1).
In this example, $h_i(x)=1+xu_i$ where $u_i$ is the $i$'th generator.
As in the case of the symmetric group (cf. \cite{KF1}),
we define the $B_n$-analogues of the Schubert polynomials
by expanding corresponding expression in the basis of group elements:
$$ \bn(x_1,\dots,x_n)=\sum_{w\in W^{(n)}} \bn_w(x_1,\dots,x_n)\ w\ ;
\tag{6.1} $$
similarly, the corresponding symmetric functions (stable Schubert
polynomials) are defined by
$$ \hn(x_1,\dots,x_k) = \sum_{w\in W^{(n)}} \hn_w(x_1,\dots,x_k)\ w\ .
\tag{6.2} $$

These definitions can be straightforwardly restated in terms of
reduced decompositions and ``compatible sequences''
(cf. \cite{BJS, FS}).
Use (4.1)-(4.2) to rewrite (6.2) as
$$ \hn_w(x_1,\dots,x_k)=
\sum_{a_1\dots a_l\in R(w)}\
\sum\Sb 1\leq b_1\leq\dots\leq b_l\leq k \\
a_i<a_{i+1}>a_{i+2}\ \Longrightarrow\ b_i<b_{i+2} \endSb
2^{\gamma(\bold a,\bold b)} x_{b_1} x_{b_2} \cdot\cdot\cdot x_{b_l}
\tag{6.3}
$$
where $R(w)$ is the set of reduced decompositions of $w$
and
$$\gamma(\bold a,\bold b)=\#\{b_i\} - \#\{i\ :\ a_i=0\}$$
(here $\#\{b_i\}$ denotes the number of {\it different} entries
in the sequence $b_1$,\dots,$b_l$).
Correspondingly, (5.1) can be presented as
$$ \bn_w(x_1,\dots,x_n)=
\sum\Sb uv=w \\
l(u)+l(v)=l(w) \\
v\in A_{n-1} \endSb
\hn_u(x_1,\dots,x_n) \ss_v(-x_1,\dots,-x_{n-1})
\tag{6.4}
$$
where, as before, $\ss_v$ is the ordinary Schubert polynomial
for $A_{n-1}=S_n$.
It is also possible to entirely rewrite in terms of reduced
decompositions and compatible sequences the definition
of Theorem 5.1. We avoid doing that since the resulting formulas
are rather messy; we also think that the following ``geometric''
approach (cf. \cite{FK1, Sec. 6}) is more natural.

Both $\bn_w$ and $\hn_w$ have a direct combinatorial interpretation
in terms of ``resolved configurations'';
this interpretation can actually be applied to any polynomials
which come from any configuration $\cc$.
Take all the intersection points of $\cc$ and ``resolve'' each of
them either as $\times$ or as $\asymp$~.
Then take all points of reflection and resolve each of them
either as $\smile$ or as $\vee$~; the latter corresponds
to changing a ``sign'', or a ``spin'', of corresponding string.
If a configuration has $N$ intersection and reflection points
altogether, then there are $2^N$ ways of producing such a resolution.
Each of the $2^N$ resolved configurations is a ``signed braid''
which naturally gives an element $w$ of the hyperoctahedral group.
Reading the $\times$- and $\vee$-points from left to right
produces a decomposition of $w$ into a product of generators.
Let $\cc_w$, for a given $w$, denote the set of resolved configurations
which give $w$ and for which this decomposition is {\it reduced}.
Then the polynomials $\Phi_w$ associated with $\cc$
(that is,
$$ \Phi(\cc; x_1,x_2,\dots) = \sum_w \Phi_w(x_1,x_2,\dots)\ w\quad ) $$
can be expressed as
$$ \Phi_w(x_1,x_2,\dots) =
\sum_{c\in\cc_w} \Bigl(\Bigl( \prod (x_i-x_j)
\Bigr)\Bigl(\prod x_k\Bigr)\Bigr)
\tag{6.5} $$
where the first product is taken over all intersections in $c$
and the second one --- over all ``change-sign'' (i.e., $\vee$-)
points.

This interpretation enables us to prove some {\it stability}
properties of $\bn_w$ and $\hn_w$.

\proclaim{6.1 Theorem}
Let $\wn$ and $\wm$, $n<m$, be the hyperoctahedral groups
with generators $s_1$,\dots,$s_n$ and $s_1$,\dots,$s_m$,
respectively.
Then, for any $w\in\wn\subset\wm$,
$$ \bn_w(x_1,\dots,x_n)=\bm_w(x_1,\dots,x_n,\underbrace{0,\dots,0}_{m-n})
\tag{6.6}
$$
and
$$ \hn_w(x_1,\dots,x_k)=\hm_w(x_1,\dots,x_k)\ .
\tag{6.7}
$$
\endproclaim

In other words, $\hn_w$~'s are stable (so we may drop superscripts)
and $\bn_w$~'s are not --- but their coefficients are.
Indeed, (6.6) means that the coefficient of any monomial in
$\bn_w(x_1,\dots,x_n)$ stabilizes as $n\to\infty$.
This allows to introduce well-defined formal power series
$$ \bb_w(x_1,x_2,\dots) = \lim_{n\to\infty}\bn_w(x_1,\dots,x_n)
\tag{6.8} $$
which could be viewed as a stable $B_n$-analogue of ordinary
Schubert polynomials.

{\bf Proof of Theorem 6.1}
It suffices to prove the case $m=n+1$;
the general statement follows by induction.
Let $\cc^{(n+1)}$ be the defining configuration for
$\bb^{(n+1)}$ or $H^{(n+1)}$.
Since $w\in \wn$ and the words in $\cc^{(n+1)}_w$ are reduced,
none of them may contain the last generator $s_n$.
In other words, all the intersections at the uppermost level
of $\cc^{(n+1)}$ have to be resolved as $\asymp$.
Then the upper string of the resulting braid does not contribute
anything (cf. (6.5)); thus we get the same result as if
we started with the configuration obtained from $\cc^{(n+1)}$
by taking out its upper boundary.
In the case of $H^{(n+1)}$, this modified configuration
is exactly the one which produces $\hn$, and (6.7) follows.
In the case of $\bb^{(n+1)}$, the modified configuration,
with $x_n=0$, is given on Figure 9 where it is transformed
to the one of $\bn$ by taking out the segment that meets
other ones at zero ``angles'' only.\ \qed

\medskip

\noindent {\bf Divided differences.}
Recall that the divided difference operator $\partial_i$
is defined by
$$ \partial_i f(x_1,\dots) =
\frac{fx_1,\dots) - f(\dots,x_{i+1},x_i,\dots)}{x_i-x_{i+1}}\ ;
\tag{6.9} $$
note that the denominator might well be $x_{i+1}-x_i$ that would
suit us better in what follows;
however, we keep the standard notation.

\proclaim{6.2 Theorem}
For any $w\in\wn$ and $i\geq 1$,
$$ -\partial_i\bn_{ws_i}\ =\ \cases
\bn_w\quad \text{\rm if} \quad l(ws_i)=l(w)+1 \\
0\qquad \text{\rm otherwise}
\endcases
\tag{6.10} $$
\endproclaim

(Unfortunately, (6.10) is {\it false} for $i=0$.)

{\bf Proof.}
As before, let $u_i$ be generators of the nilCoxeter algebra.
Then the theorem is equivalent to the identity
$$ -\partial_i\ \sum_w \bn_{ws_i}ws_i
= \sum_w \bn_w wu_i $$
which, in turn, can be rewritten as
$$ -\partial_i\bn = \bn u_i\ ; $$
the latter follows from (5.1) and the identity
$\partial_i\ss=\ss u_i$ (see \cite{FS, Lemma 3.5})
which is just another way of stating the divided-differences
recurrence for the ordinary Schubert polynomials.\ \qed

Note that a result analogous to Theorem 6.2 holds for the power
series $\bb_w$ defined in (6.8).


\ssectitle{7. Computations}

Polynomials $\bn_w$ can be computed by expanding (5.4)
in the basis of permutations.
One can also use (6.3)-(6.4) and/or the recurrences (6.10).
We give below the results of our computations for $n=1$, 2,
and 3; we used the formulas of Example 5.2 and then double-checked
the results with (6.10).

\bigskip
\bigskip

\noindent $\boxed{n=1}$ \qquad $\bb^{(1)}(x_1) = h_0(x_1) = 1+x_1u_0$
\medskip
{\settabs 4 \columns
\+ $w$	&  	 $\bb^{(1)}_w$ \cr
\+ \cr
\+ 1	 	& 1 \hfill \cr
\+ $u_0$ \hfill	&  $x_1$ \hfill \cr
}

\newpage

\noindent$\boxed{n=2}$
$$
\align
&\bb^{(2)}(x_1,x_2) \\
&=h_1(x_2) h_0(x_2) h_1(x_1+x_2) h_0(x_1)
=(1+x_2u_1)(1+x_2u_0)(1+(x_1+x_2)u_1)(1+x_1u_0)
\endalign
$$
\medskip
{\settabs 4 \columns
\+ $w$	&  	 $\bb^{(2)}_w$ \cr
\+ \cr
\+ 1	 	& 1 \hfill \cr
\+ $u_0$ \hfill	&  $x_1+x_2$ \hfill \cr
\+ $u_1$ \hfill	&  $x_1+2x_2$ \hfill \cr
\+ $u_0u_1$ \hfill	&  $x_2(x_1+x_2)$ \hfill \cr
\+ $u_1u_0$ \hfill	&  $(x_1+x_2)^2$ \hfill \cr
\+ $u_0u_1u_0$ \hfill	&  $x_1x_2(x_1+x_2)$ \hfill \cr
\+ $u_1u_0u_1$ \hfill	&  $x_2^2(x_1+x_2)$ \hfill \cr
\+ $w_0$ \hfill	&  $x_1x_2^2(x_1+x_2)$ \hfill \cr
}

\bigskip
\bigskip

\noindent $\boxed{n=3}$
$$\align
\bb^{(3)}(x_1,x_2) &= h_2(x_3) h_1(x_3) h_2(x_2) h_0(x_3)
h_1(x_2+x_3) h_2(x_1+x_3) h_0(x_2) h_1(x_1+x_2) h_0(x_1) \\
&= (1+x_3u_2)(1+x_3u_1)(1+x_2u_2)(1+x_3u_0)(1+(x_2+x_3)u_1) \\
&\qquad\qquad\times
(1+(x_1+x_3)u_2)(1+x_2u_0)(1+(x_1+x_2)u_1)(1+x_1u_0)
\endalign
$$

\noindent In the following table, \quad $L=x_1+x_2+x_3$,
\quad $K=(x_1+x_2)(x_1+x_3)(x_2+x_3)$.
\bigskip
{\settabs 12 \columns
\+ $w$	& & & \ \ $\#(R(w))$ & 	& $\bb^{(3)}_w$ \cr
\+ \cr
\+ 123 \ \ 1	& & & & 1 	& 1 \hfill \cr
\+ ${\overline 1}$23 \ \ $u_0$ \hfill	& & & & 1 	& $L$ \hfill \cr
\+ 213 \ \ $u_1$ \hfill	& & & & 1 	& $x_1+2x_2+2x_3$ \hfill \cr
\+ 132 \ \ $u_2$ \hfill	& & & & 1 	& $x_1+x_2+2x_3$ \hfill \cr
\+ 2${\overline 1}$3 \ \ $u_0u_1$ \hfill	& & & & 1 	& $(x_2+x_3)L$ \hfill \cr
\+ ${\overline 1}$32 \ \ $u_0u_2$ \hfill	& & & & 2 	& $(x_1+x_2+2x_3)L$ \hfill
\cr
\+ ${\overline 2}$13 \ \ $u_1u_0$ \hfill	& & & & 1 	& $L^2$ \hfill \cr
\+ 231 \ \ $u_1u_2$ \hfill	& & & & 1 	& $2x_3L+x_1x_2$ \hfill \cr
\+ 312 \ \ $u_2u_1$ \hfill	& & & & 1 	& $L^2+(x_2+x_3)^2$ \hfill \cr
\+ ${\overline 2}{\overline 1}$3 \ \ $u_0u_1u_0$ \hfill	& & & & 1 	& $K$ \hfill
\cr
\+ 23${\overline 1}$ \ \ $u_0u_1u_2$ \hfill	& & & & 1 	&
$x_3(x_1+x_3)(x_2+x_3)$ \hfill \cr
\+ 3${\overline 1}$2 \ \ $u_0u_2u_1$ \hfill	& & & & 2 	&
$(x_2+x_3)(L^2+x_2x_3)$ \hfill \cr
\+ 1${\overline 2}$3 \ \ $u_1u_0u_1$ \hfill	& & & & 1 	&
$(x_2+x_3)(x_3L+x_2^2+x_1x_2)$ \hfill \cr
\+ ${\overline 2}$31 \ \ $u_1u_0u_2$ \hfill	& & & & 2 	& $x_3L^2+K$ \hfill \cr
\+ 321 \ \ $u_1u_2u_1$ \hfill	& & & & 2 	& $K+x_1x_2^2+x_3(x_1+3x_2+2x_3)L$
\hfill \cr
\+ ${\overline 3}$12 \ \ $u_2u_1u_0$ \hfill	& & & & 1 	& $L(x_1^2+x_2^2+x_3^2+
x_1x_2+x_1x_3+x_2x_3)+x_1x_2x_3$ \hfill \cr
\+ ${\overline 1}{\overline 2}$3 \ \ $u_0u_1u_0u_1$ \hfill	& & & & 2 	&
$(x_2+x_3)K$ \hfill \cr
\+ ${\overline 2}$3${\overline 1}$ \ \ $u_0u_1u_0u_2$ \hfill	& & & & 2 	&
$x_3K$ \hfill \cr
\+ 32${\overline 1}$ \ \ $u_0u_1u_2u_1$ \hfill	& & & & 3 	&
$x_3(x_2+x_3)(x_1+x_3)
(x_1+2x_2+x_3)$ \hfill \cr
\+ ${\overline 3}{\overline 1}$2 \ \ $u_0u_2u_1u_0$ \hfill	& & & & 2 	& $KL$
\hfill \cr
\+ 13${\overline 2}$ \ \ $u_1u_0u_1u_2$ \hfill	& & & & 1 	&
$x_3^2(x_2+x_3)(x_1+x_3)$ \cr
\+ 3${\overline 2}$1 \ \ $u_1u_0u_2u_1$ \hfill	& & & & 2 	& $(x_2+x_3)K+
x_2x_3(x_2+x_3)L$ \hfill \cr
\+ ${\overline 3}$21 \ \ $u_1u_2u_1u_0$ \hfill	& & & & 3 	&
$x_3^3L+KL+x_3(x_1+x_2)(L^2-x_1x_2)$ \hfill \cr
\+ 1${\overline 3}$2 \ \ $u_2u_1u_0u_1$ \hfill	& & & & 1 	& $(x_2+x_3)
(x_3^2L+x_2^2L+x_1x_2x_3)$ \hfill \cr
\+ ${\overline 1}$3${\overline 2}$ \ \ $u_0u_1u_0u_1u_2$ \hfill	& & & & 3 	&
$x_3^2K$ \hfill \cr
\+ 3${\overline 2}{\overline 1}$ \ \ $u_0u_1u_0u_2u_1$ \hfill	& & & & 2 	&
$x_2x_3K$ \hfill \cr
\+ ${\overline 3}$2${\overline 1}$ \ \ $u_0u_1u_2u_1u_0$ \hfill	& & & & 5 	&
$x_3KL$ \hfill \cr
\+ ${\overline 1}{\overline 3}$2 \ \ $u_0u_2u_1u_0u_1$ \hfill	& & & & 3 	&
$K(x_2^2+x_2x_3+x_3^2)$ \hfill \cr
\+ 31${\overline 2}$ \ \ $u_1u_0u_1u_2u_1$ \hfill	& & & & 3 	& $x_3^2
(x_1+2x_2)(x_1+x_3)(x_2+x_3)$ \hfill \cr
\+ ${\overline 3}{\overline 2}$1 \ \ $u_1u_0u_2u_1u_0$ \hfill	& & & & 2 	&
$K(x_1x_2+x_1x_3+x_2x_3)$ \hfill \cr
\+ 2${\overline 3}$1 \ \ $u_1u_2u_1u_0u_1$ \hfill	& & & & 3 	&
$x_2x_3(x_2+x_3)(x_3L+x_2^2+x_1x_2)+
K(x_2^2+x_2x_3+x_3^2)$ \hfill \cr
\+ 12${\overline 3}$ \ \ $u_2u_1u_0u_1u_2$ \hfill	& & & & 1 	&
$x_3^3(x_2+x_3)(x_1+x_3)$ \cr
\+ 3${\overline 1}{\overline 2}$ \ \ $u_0u_1u_0u_1u_2u_1$ \hfill	& & & & 5 	&
$x_2x_3^2K$ \hfill \cr
\+ ${\overline 3}{\overline 2}{\overline 1}$ \ \ $u_0u_1u_0u_2u_1u_0$ \hfill	&
& & & 2 	& $x_1x_2x_3K$ \hfill \cr
\+ 2${\overline 3}{\overline 1}$ \ \ $u_0u_1u_2u_1u_0u_1$ \hfill	& & & & 5 	&
$x_2x_3(x_2+x_3)K$ \hfill \cr
\+ ${\overline 1}$2${\overline 3}$ \ \ $u_0u_2u_1u_0u_1u_2$ \hfill	& & & & 4 	&
$x_3^3K$ \hfill \cr
\+ ${\overline 3}$1${\overline 2}$ \ \ $u_1u_0u_1u_2u_1u_0$ \hfill	& & & & 5 	&
$x_3^2(x_1+x_2)K$ \hfill \cr
\+ ${\overline 2}{\overline 3}$1 \ \ $u_1u_0u_2u_1u_0u_1$ \hfill	& & & & 5 	&
$(x_2x_3L+x_1x_2^2+x_1x_3^2)K$ \hfill \cr
\+ 21${\overline 3}$ \ \ $u_1u_2u_1u_0u_1u_2$ \hfill	& & & & 4 	& $x_3^3
(x_1+2x_2)(x_1+x_3)(x_2+x_3)$ \hfill \cr
\+ ${\overline 3}{\overline 1}{\overline 2}$ \ \ $u_0u_1u_0u_1u_2u_1u_0$
\hfill	&  & &  & 7 	& $x_1x_2x_3^2K$ \hfill \cr
\+ ${\overline 2}{\overline 3}{\overline 1}$ \ \ $u_0u_1u_0u_2u_1u_0u_1$
\hfill	& & & & 7 	&
$x_1x_2x_3(x_2+x_3)K$ \hfill \cr
\+ 2${\overline 1}{\overline 3}$ \ \ $u_0u_1u_2u_1u_0u_1u_2$ \hfill	& & & & 9
& $x_2x_3^3K$ \hfill \cr
\+ 1${\overline 3}{\overline 2}$ \ \ $u_1u_0u_1u_2u_1u_0u_1$ \hfill	& & & & 5
& $x_2^2x_3^2K$ \hfill \cr
\+ ${\overline 2}$1${\overline 3}$ \ \ $u_1u_0u_2u_1u_0u_1u_2$ \hfill	& & & & 9
	& $(x_1+x_2)x_3^3K$ \hfill \cr
\+ ${\overline 1}{\overline 3}{\overline 2}$ \ \ $u_0u_1u_0u_1u_2u_1u_0u_1$
\hfill	& & & & 12 	& $x_1x_2^2x_3^2K$ \hfill \cr
\+ ${\overline 2}{\overline 1}{\overline 3}$ \ \ $u_0u_1u_0u_2u_1u_0u_1u_2$
\hfill	& & & & 16 	& $x_1x_2x_3^3K$ \hfill \cr
\+ 1${\overline 2}{\overline 3}$ \ \ $u_1u_0u_1u_2u_1u_0u_1u_2$
 \hfill
& & & & 14	& $x_2^2x_3^3K$ \hfill \cr
\+ ${\overline 1}{\overline 2}{\overline 3}$ \ \ $u_0u_1u_0u_1u_2u_1u_0u_1u_2$
 \hfill
& & & & 42	& $x_1x_2^2x_3^3K$ \hfill \cr
}

\ssectitle{8. Symmetric functions}

This section is devoted to studying the basic properties of
the symmetric functions $H_n$ defined by (6.2), (4.2), and (4.1).

In the nilCoxeter case, Theorem 4.4 immediately
allows to establish the
following connection between $H_w$'s and Stanley's symmetric
functions \nopagebreak{(stable Schubert polynomials)~$G_w$.}

\proclaim{8.1 Corollary}
Let $\xx=(x_1,\dots,x_n)$.
Let $w$ be an element of the symmetric subgroup
of the hyperoctahedral group $\wn$ that is generated by
$s_1$, \dots, $s_{n-1}$.
Then
$$ \hn_w(\xx) = G_w^{\text{super}}(\xx,\xx) $$
where $G^{\text{super}}_w$ is the canonical superfication
of the stable Schubert polynomial $G_w$.\ \qed
\endproclaim

The last formula implies that, for such $w$,
$H_w$ is a nonnegative integer linear combination of
Schur $P$-functions.
Also, the definition of the functions $H_w$ implies that
they are, for any $w$, {\it some} linear combinations of $P$-functions.
Tao Kai Lam recently found a proof that, in fact,
$H_w$ is always a {\it nonnegative integer} linear combination of
$P$-functions.

It can be shown that the [skew] $P$-functions
themselves are a special case of $H_w$'s.
To do that, we generalize, in a more or less straightforward
way, the corresponding $S_n$-statement about
321-avoiding permutations (see \cite{BJS}).

\proclaim{8.2 Theorem}
Let $\sigma$ be a skew shifted shape presented in a standard ``English''
notation (see, e.g., \cite{SS}).
Define a ``content'' of each cell of $\sigma$ to be the difference
between the number of column and the number of row which this cell is in.
(E.g., the content of a cell lying on the main diagonal is 0.)
Read the contents of the cells of $\sigma$ column by column,
from top to bottom; this gives a sequence $a_1$, \dots, $a_l$.
Define an element $w_\sigma$ of the hyperoctahedral group $\wn$
by $w_\sigma=s_{a_1}\cdot\cdot\cdot s_{a_l}$
(in fact, $a_1$, \dots, $a_l$ is a reduced decomposition of $\sigma$).
Then $H_{w_\sigma}=P_\sigma$ where $P_\sigma$ is the skew Schur
$P$-function corresponding to the shape $\sigma$.
\endproclaim

One could also ask: which elements $w\in\wn$ can be presented as
$w_\sigma$ (see Theorem~8.2)?
The answer can be presented
in either of the following equivalent ways
(both are informal though unambigous):

\noindent (i) those $w$ for which the set of reduced decompositions
$R(w)$ contains a single commutation class (i.e, they can be obtained from
each other by switches of commuting variables, without any use of Coxeter
relations);

\noindent (ii) those $w$ which avoid the following patterns:
$$ 3\ 2\ 1\qquad \overline{3}\ 2\ 1\qquad 3\ 2\ \overline{1}\qquad
\overline{3}\ 2\ \overline{1}
\qquad 1\ \overline{2}\qquad \overline{1}\ \overline{2}$$
where $\overline{i}$ denotes the element $i$ of a signed permutation
that has changed its sign.

Technical proofs of the last statement and of Theorem 8.2
will be given elsewhere.

\noindent {\bf $H_w$ as a limit of $\bb_w$.}
Similarly to the $S_n$ case, the symmetric functions $\hn_w$
can be obtained as some kind of limit of $\bm_w$ as $m\to\infty$.

\proclaim{8.3 Theorem}
Let $w$ be any element of the hyperoctahedral group
with $n$ generators.
Then
$$ \lim_{N\to\infty} \bb_w^{(n+N)} (\underbrace{0,\dots,0}_N,
x_1,\dots,x_n)\ =\ \hn_w(x_1,\dots,x_n)\ .$$
Even more: if $N\geq n-1$, then
$$ \bb_w^{(n+N)} (0,\dots,0,
x_1,\dots,x_n)\ =\ \hn_w(x_1,\dots,x_n)\ .$$
\endproclaim

{\bf Proof.} Similar to the proof of Theorem 6.1.\ \qed

\ssectitle{References}

\item{\quad {[BJS]}\quad}
S.C.Billey, W.Jockush, and R.P.Stanley, Some combinatorial properties
of Schubert polynomials, manuscript, MIT, 1992.

\item{\quad {[Ch]}\quad}
I.Cherednik, Notes on affine Hecke algebras.I,
{\it Max-Planck-Institut Preprint} MPI/91-14\rm , 1991.

\item{\quad {[FK1]}\quad}
S.Fomin, A.N.Kirillov, The Yang-Baxter equation, symmetric functions,
and Schubert polynomials, to appear in {\it Proceedings of the 5th
International Conference on Formal Power Series and Algebraic Combinatorics},
Firenze, 1993.

\item{\quad {[FK2]}\quad}
S.Fomin, A.N.Kirillov, Grothendieck polynomials and the Yang-Baxter equation,
manuscript, 1993.

\item{\quad {[FK3]}\quad}
S.Fomin, A.N.Kirillov, Universal exponential solution
of the Yang-Baxter equation,
manuscript, 1993.

\item{\quad {[FS]}\quad}
S.Fomin, R.P.Stanley, Schubert polynomials and the nilCoxeter algebra,
{\it Advances in Math.}, to appear; see also
{\it Report No.18 {\rm (1991/92),} Institut Mittag-Leffler},
1992.

\item{\quad {[L]}\quad}
A.Lascoux, Polyn\^omes de Schubert. Une approche historique,
{\it S\'eries formelles et combinatoire alg\'ebri\-que},
P.Leroux and C.Reutenauer, Ed., Montr\'eal, LACIM, UQAM, 1992, 283-296.

\item{\quad {[M]}\quad}
I. Macdonald, {\it Notes on Schubert polynomials},
Laboratoire de combinatoire et d'in\-for\-ma\-ti\-que math\'ema\-tique
(LACIM), Universit\'e du Qu\'ebec \`a Montr\'eal, Montr\'eal, 1991.

\item{\quad {[SS]}\quad}
B.E.Sagan, R.P.Stanley, Robinson-Schensted algorithms
for skew tableaux,
{\it J.Com\-bin. Theory}, Ser.A  {\bf  55} (1990), 161-193.

\item{\quad {[S]}\quad}
R.P.Stanley,
On the number of reduced decompositions of elements of Coxeter groups,
\it European J. Combin. \bf 5 \rm (1984), 359-372.

\enddocument